\documentclass[aps,twocolumn,prl,showpacs]{revtex4}

\usepackage{amsfonts}
\usepackage{amsmath}
\usepackage{amssymb}
\usepackage{graphicx}

\newcommand{\Zeff}{Z_{\text{eff}}}
\newcommand{\keff}{\kappa_{\text{eff}}}

\newcommand{\ueff}{u_{\text{eff}}}

\newcommand{\tildeZeff}{\tilde{Z}_{\text{eff}}}

\begin{document}

\title{Density dependent interactions and structure of charged colloidal dispersions in the weak screening regime}

\author{L. F. Rojas-Ochoa$^{(1,2)}$}
\author{R. Casta\~{n}eda-Priego$^{(3)}$}
\author{V. Lobaskin$^{(4)}$}\altaffiliation[Present address: ]{School of Physics, University College Dublin, Belfield, Dublin 4, Ireland.}
\author{A. Stradner$^{(2)}$}
\author{F. Scheffold$^{(2)}$}
\author{P. Schurtenberger$^{(2)}$}

\affiliation{$^{(1)}$Departamento de F\'{i}sica, Cinvestav-IPN, Av.
Instituto Polit\'{e}cnico Nacional 2508, 07360 M\'{e}xico D. F.,
Mexico}
\affiliation{$^{(2)}$Physics Department, University of
Fribourg, Chemin de Mus\'{e}e 3, 1700 Fribourg, Switzerland}
\affiliation{$^{(3)}$Instituto de F\'{i}sica, Universidad de
Guanajuato, Loma del Bosque 103, 37150 Le\'{o}n, Mexico}
\affiliation{$^{(4)}$Physik-Department, Technische Universit\"at
M\"unchen, James-Franck-Str., D-85747 Garching, Germany}

\date{\today}


\begin{abstract}

We determine the structure of charge-stabilized colloidal suspensions at low ionic strength over an extended range of particle volume fractions using a combination of light and small angle neutron scattering experiments. The variation of the structure factor with concentration is analyzed within a
one-component model of a colloidal suspension. We show that the observed structural behavior corresponds to a non-monotonic density dependence of the colloid effective charge and the mean interparticle interaction energy. Our findings are corroborated by similar observations from primitive model computer simulations of salt-free colloidal suspensions.

\end{abstract}

\maketitle

Deionized colloidal dispersions have recently attracted much experimental \cite{Palberg2004,jcppalberg06,Haro-Perez2006} and theoretical attention
\cite{Trizac2004,mjellium,Trizac2007,Lobaskin2007} because of their unique role as simple model systems in the field of surface charge regulation. In addition to experimental challenges posed by these systems, such as the need of control over the ionic strength of the medium, a number of intriguing observations have been made that challenge classical theories of electrostatic screening and electrokinetics (e.g., see \cite{Palberg2004,Royall2006,Tata2006} and references therein). Some of the conceptual difficulties in the understanding of these systems stem from the counterion dominance in the ionic cloud of the colloid, which makes the
screening ability of the medium dependent on the colloidal concentration. At the same time, the counterion condensation, which is controlled by the balance of the counterion chemical potential between the bulk suspension and the condensed layer, would depend on the concentration through the state-dependent screening parameters \cite{jellium1984,PBcell1984,Trizac2004}. Therefore, modeling these systems within the usual one-component description can only be achieved using effective density-dependent interaction parameters, charge and screening length \cite{Belloni2000}. The dependence of the colloid effective charge on the volume fraction \cite{Gisler1994,Trizac2004,mjellium} or background ionic strength \cite{Netz2003} has been predicted in a few mean-field models and is expected to be non-monotonic. However, this non-monotonic behavior has never been validated experimentally.

In this Letter, we present a study of the density dependence of the colloidal structure and interactions at low ionic strength. Experiments have been carried out over a wider range of concentrations than in any previous work. We analyze the particle structure factor obtained from light and neutron scattering at different particle volume fractions and demonstrate that (i) the variation of colloidal structure in deionized suspensions can be interpreted in terms of a one-component model (OCM) that considers Yukawa-like pair interactions between colloids with state-dependent parameters and, (ii) the non-monotonic concentration dependence of the interaction energy, as
characterized by the colloid effective charge, is a general feature of deionized colloidal dispersions. Primitive model (PM) numerical simulations of salt-free suspensions corroborate our observations.
\begin{table}[b]
    \centering
     \caption{Polystyrene latex spheres used in the experiments. Bjerrum length, $\lambda_{\text{B}} =  e^{2} / \left( 4 \pi \, \epsilon \, \epsilon_{0} \, k_{\text{B}} \, T \right) $, from the dielectric permittivity data in Ref.~\cite{ethanol}.}
    \vspace{0.1cm}
    \begin{tabular}{c c c c c}
    \hline\hline
    Sample $\#$ & Radius & Bjerrum       & Volume          & Scattering \\
                & ($nm$) & length ($nm$) & fraction ($\%$) & Technique\\
    \hline
    S1 &  54.9 $\pm$ 6.4 & 1.296 & 0.035 - 01.15 & 3D-DLS \\
    S2 &  58.7 $\pm$ 8.6 & 1.383 & 0.135 - 01.48 & 3D-DLS \\
    S3 &  54.7 $\pm$ 5.3 & 1.482 & 2.500 - 15.50 & SANS \\
    \hline\hline
    \end{tabular}
    \label{samples}
\end{table}

Stock suspensions of negatively charged particles (sulfonate polystyrene latex) were purchased from  Interfacial Dynamics Corporation (Portland, USA) and master dispersions were prepared following the protocol described in
Ref.~\cite{LuisEPL2002,LuisPRL2004}. The samples were prepared by diluting the master suspensions directly into quartz cells with water/ethanol mixtures; filtered several times (pore size $0.2 \, \mu m$) to remove dust particles. These mixtures were chosen in order to avoid crystallization by modifying the solvent dielectric constant \cite{ethanol}, thereby providing a well controlled model system of charged dispersions with volume fractions as high as $\sim 16 \%$. Accurate values of particle radii and polydispersities, as obtained from DLS and SANS experiments, are given in Table~\ref{samples}.

The effective interparticle interaction can be quantified through the analysis of the suspension static structure. This requires probing length scales comparable to the size of the particles. Usually such information can be obtained from static and dynamic light scattering (SLS/DLS) in the single scattering regime \cite{Peter1993}. In dense suspensions, however, these techniques cannot be applied in most cases due to multiple scattering of
light. An elegant way to overcome this limitation is the use of modern scattering techniques such as 3D dynamic light scattering (3D-DLS) \cite{3DDLS} or small angle neutron scattering (SANS). While 3D-DLS provides valuable information at small and intermediate values of the scattering vector $q$, access to large $q$-vectors from SANS allows to normalize the static structure factor $S(q)$ and, in that way, unambiguously determine the suspension structure. A detailed description of the light scattering (LS) experiments is given in Ref.~\cite{LuisEPL2002}.

Multiple scattering of neutrons in our SANS experiments has been suppressed by partially contrast matching the particles using H$_{2}$O/ethanol -- D$_{2}$O/deuterated ethanol mixtures. The samples were kept in stoppered quartz cells (Hellma, Germany) with a path length of $2 \, mm$ and containing a mixed-bed of ion-exchanger resins; deionization was typically completed within $2$ weeks. The measurements were performed with a mean neutron
wavelength of $1.27 \, nm$ and at a detector distance of $20.3 \, m$, which corresponds to a $q$-range of $0.01$-$0.1 \, nm^{-1}$. For details on the initial data treatment and data analysis see Ref.~\cite{LuisPRE2002} and references therein. The SANS scattering intensity can be expressed as $I(q) = A \, P(q)_{\text{SANS}} \, S(q)_{\text{SANS}}$, where $P(q)_{\text{SANS}}$ is the normalized effective particle form factor, $S(q)_{\text{SANS}}$ is the effective interparticle structure factor and $A$ is an amplitude, which is proportional to the particle number density, $n$, and to the square of the neutron scattering length contrast between the particles and background fluid \cite{Wu1987}. The effective SANS static structure factor can be determined from $S(q)_{\text{SANS}} = ( A_{0} / A ) \, [ I(q) / I_{0}(q) ]$, where $I_{0}(q) = A_{0} \, P(q)_{\text{SANS}}$ is the intensity scattered by a suspension of non-interacting particles with amplitude $A_{0}$.

On the theory side, the static structure factor of colloidal dispersions can be calculated by solving the Ornstein-Zernike (OZ) equation together with the Rogers-Young (RY) closure relation \cite{DAguanno1992} for particles interacting with an effective pair potential $\ueff(r)$ of the Yukawa form
\begin{equation}\label{eq1}
   \beta \ueff(r) = \Zeff^{2} \, \lambda_{\text{B}} \left[ \frac{\exp(\keff \, a)}
   {1 + \keff \, a} \right]^2 \frac{\exp(-\keff \, r)}{r} \, ,
\end{equation}
with an effective charge $\Zeff$, colloid radius $a$, thermal energy $\beta^{-1} = k_{\text{B}} \, T$, screening parameter $\kappa^{2}_{\text{eff}} = 4 \pi \, \lambda_{\text{B}} \, (\Zeff \, n + n_{\text{s}})$ and concentration of salt ions $n_{\text{s}} = 2 \times 1000 \, N_\text{A} c_\text{s}$, where $c_\text{s}$ is the molar concentration \cite{Belloni2000}. This procedure has been successfully used for calculating the structure in the strong screening regime at $\keff \, a \gg 1$ \cite{DAguanno1992} but it becomes less accurate in systems with thick double layers as the interparticle potential deviates from the Yukawa shape \cite{Lobaskin2003a}. Despite this deficiency, an appropriate choice of $\Zeff$ still allows one to predict the structure factor of strongly interacting systems over a wide range of conditions as long as the pair interaction energy at the mean interparticle
distance is correctly reproduced \cite{lobaskin.v:2001}. Nevertheless, an agreement of the OCM structure and the mean interaction energy with the original ones does not guarantee that other thermodynamic properties, such as the osmotic pressure, can be predicted using the effective parameters. The difficulty with the pressure originates from the loss of the ionic degrees of
freedom in the coarse-grained description. Various solutions to this issue have been suggested recently. In one, the missing pressure contributions have been recovered via inclusion of the boundary terms \cite{Trizac2007}. Alternatively, one can use a modified OZ-RY approach to get the effective parameters in the fit procedure \cite{mjellium}. The latter replaces the OCM (Yukawa fluid) pressure by the osmotic pressure known \emph{a priori} from primitive model simulations or mean-field models (e.g. jellium) in calculating the isothermal compressibility. We use this route to extract the effective charges and effective potentials from the structure data measured in experiment and simulations.
%
%
\begin{figure}[b]
\centering
\includegraphics[width=210.0pt]{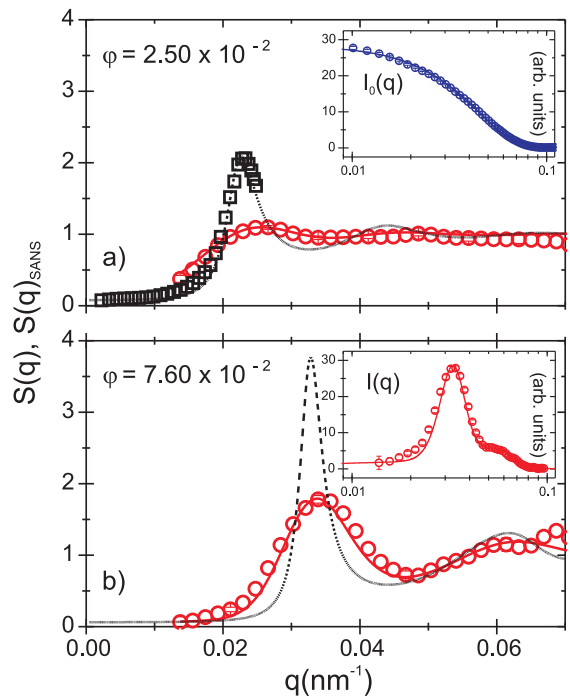}
\caption{(color online) Normalized scattering intensities of samples a) $\varphi = 0.025$ and, b) $\varphi = 0.076$. The squares are from 3D-DLS and the circles from SANS. The dashed line is the calculated $S(q)$ and the continuous line its SANS experimental convolution, $S(q)_{\text{SANS}}$. Insets: scattering intensities for a) a diluted screened suspension and, b) $\varphi = 0.076$.}\label{fig1}
\end{figure}
%
%

Fig.~\ref{fig1} displays the normalized scattering intensity for two of the samples S3 in Table~\ref{samples}: a) $\varphi = 2.5 \%$ and b) $\varphi = 7.6 \%$. In a), the squares are from 3D-DLS and circles from SANS measurements, thus illustrating the complementarity of both experimental techniques. The experimental error is estimated from the intensity statistics in a series of
measurements. For both SANS and LS, the experimental data are in good agreement with theoretical calculations (lines) based on integral equations theory using the polydisperse RY closure relation \cite{DAguanno1992} as described in Ref.~\cite{mjellium}. The background salt concentration, $c_{\text{s}} = 2 \times 10^{-7} \, \text{M}$, was determined by measuring the conductivity of our deionized solvent mixtures as in Ref. \cite{Palberg1998}.
Thus, the free parameters in our fitting procedure are the volume fraction $\varphi$ and the effective charge $\Zeff$. The dashed lines represent the calculated $S(q)$ and the continuous lines correspond to the result after its experimental convolution, $S(q)_{\text{SANS}}$, with the appropriate resolution function \cite{Barker1995}. The structure factor exhibits a pronounced maximum, whose position $q_{\text{max}}$ varies with the particle
volume fraction as $n^{-1/3}$ \cite{LuisPRL2004}. Additionally, in the Insets we also show the SANS scattering intensities for: a) a dilute, $\varphi \approx 0.005$, screened ($c_{\text{s}} \sim 0.005 \, \text{M}$ of KCl) suspension and, b) the sample $\varphi = 7.6 \%$. Here, the lines are experimentally convoluted fits with polydisperse form, $I_{0}(q) \propto P(q)_{\text{SANS}}$, and structure factor, $I(q) \propto P(q)_{\text{SANS}} \, S(q)_{\text{SANS}}$.
%
%
\begin{figure}[b]
\centering
\includegraphics[width=210.0pt]{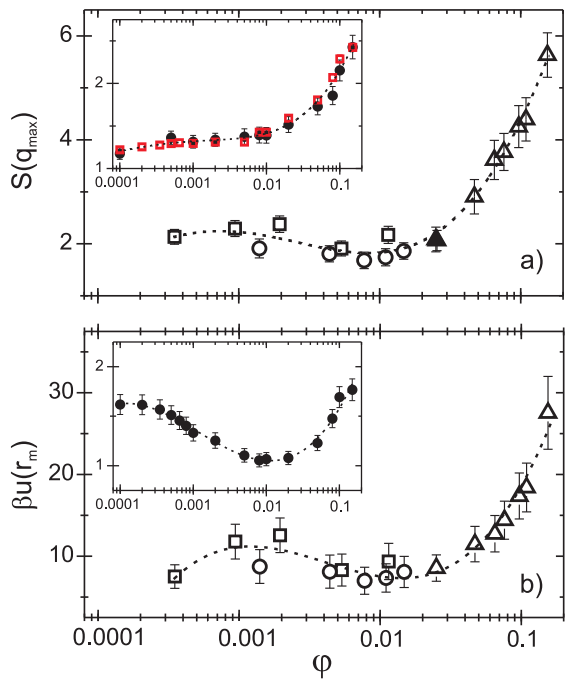}
\caption{(color online) a) $S(q_{\text{max}})$ vs. $\varphi$. Squares and circles are from 3D-DLS measurements in S1 and S2, respectively. Triangles are from 3D-DLS (solid) and SANS analysis (open) in S3. Inset: simulations from PM (solid circles) and OCM-NPT (open circles). b) The colloid interaction energies $\beta \ueff(r)$ at the mean interparticle distance $r_{\text{m}}$ for experiments and simulations (Inset). Lines are guides to the eye.}\label{fig2}
\end{figure}
%
%

In Fig.~\ref{fig2}a) we present the $\varphi$-dependence of the height of the main peak of the structure factor, $S(q_{\text{max}})$, as obtained from the LS and SANS analysis. At low volume fractions, the peak height is about 2 and varies weakly with concentration until $\varphi \approx 0.002$, then it decreases in the range $0.002 < \varphi < 0.01$ to finally grow again for $\varphi > 0.01$. We note that for $\varphi > 0.06$, the values of $S(q_{\text{max}}) > 3.0$ indicate a glassy state \cite{Glass}. While the increase of $S(q_{\text{max}})$ with concentration can be readily explained by shortening the distance between neighboring colloids and rising average interparticle repulsion, the decrease at $\varphi \approx 0.01$ is a more
complex feature. This depression can be attributed to the maximum of ionic condensation or a minimum of the effective macroion charge as discussed below. To test whether this behavior is specific to the experimental systems considered here, in the Inset of Fig.~\ref{fig2}a) we show results of PM simulations (solid symbols) of a salt-free charged colloidal suspension with charge asymmetry $200 \,$:$\, 1$. The simulations were performed for systems with $\lambda_{\text{B}} / a = 0.2$ using the same cluster Monte Carlo simulation protocol and settings as in Ref.~\cite{Lobaskin1999} with $80$ colloidal particles. The $S(q_{\text{max}})$ from simulations shows a similar trend to the experiments although the main features are less pronounced. Nevertheless, one can show that this characteristic trend of the structure factor corresponds to a quite peculiar behavior of the pair interaction potential. Fig.~\ref{fig2}b) depicts the interaction energy between nearest neighbor colloids, $\beta \ueff(r_{\text{m}})$, as extracted from fits to the structure data of systems S1-S3. $\beta \ueff(r_{\text{m}})$ follows the trend observed for $S(q_{\text{max}})$ with more pronounced features: a maximum at $\varphi \approx 0.001$, a minimum at $\varphi \approx 0.01$, and an increase at higher volume fractions. The same holds for the PM results; solid circles in the Inset of Fig.~\ref{fig2}b). Furthermore, we were able to self-consistently prove our primitive model results by using their associated
effective charges and screening lengths in OCM-NPT simulations (open squares in the Inset of Fig.~\ref{fig2}a)). Here, we used N$\, = 1000$ colloidal particles and the osmotic pressure as obtained from the PM simulations.
%
%
\begin{figure}[b]
\centering
\includegraphics[width=210.0pt]{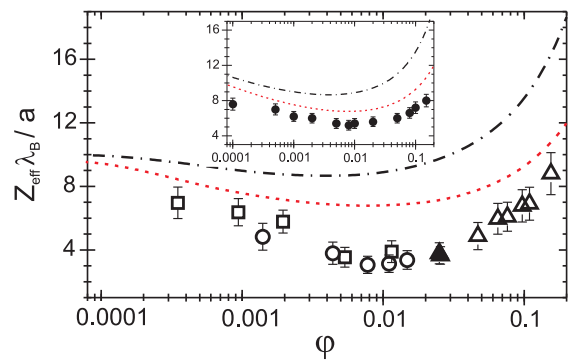}
\caption{(color online) $\tildeZeff$ vs. $\varphi$ from experiments (main) and simulations (Inset). The lines are calculations based on renormalized jellium (dashed) and PB-cell (dash-dotted) models, and considering $c_{\text{s}} = 2 \times 10^{-7} \, \text{M}$ (main) and $c_{\text{s}} = 0$ (Inset). Same
symbols as in Fig.~\ref{fig2}.}\label{fig3}
\end{figure}
%
%

The interactions between colloids are conveniently characterized by their effective charge. To compare the results between different systems we use the dimensionless ratio $\tildeZeff = \Zeff \lambda_{\text{B}} / a$ instead of the valency itself. In Fig.~\ref{fig3} we show the results of $\tildeZeff$ as obtained from fitting the scattering data with our OZ-RY scheme (same symbols as in Fig.~\ref{fig2}). Furthermore, in Fig.~\ref{fig3} we compare the data with results of the PM simulations described above (solid circles in the Inset), where $Z_{\text{bare}} \lambda_{\text{B}} / a = 40.0$ and the $\tildeZeff$ are extracted following the same protocol as in the experiments. Both sets of data show a pronounced minimum of the effective charge at $\varphi \approx 10^{-2}$.

The behavior of $\tildeZeff$ in the range $\varphi \lesssim 10^{-2}$ stems from the variation of the mean counterion concentration. Upon increasing $\varphi$, counterions in average get closer to the colloidal surface thus leading to a gradual decrease of $\tildeZeff$. At $\varphi > 10^{-2}$, in contrast, the effective charge starts increasing because the average
electrostatic potential in the bulk, or at least at $r \geq r_{\text{m}}$, becomes comparable to that on the colloid surface thus making the counterion condensation less favorable. In Fig.~\ref{fig3}, we also present results from two models, which predict a similar density dependence for $\tildeZeff$: renormalized jellium \cite{Trizac2004} and Poisson-Boltzmann cell model (PB-cell) \cite{PBcell1984}. We used $c_\text{s} = 2 \times 10^{-7} \, \text{M}$ in the experiments (main figure) and $c_\text{s} = 0$ in the simulations (Inset). These models correctly reproduce the qualitative behavior of the effective charge, although the values corresponding to the saturation regime are too high. This quantitative disagreement might be caused by various reasons. First, the best fit in our procedure corresponds to slightly lower effective charges than those in the jellium model. Moreover, the values are affected by the actual surface charge density on the colloids in experiment and ionic correlation effects in the simulations \cite{Lobaskin1999}. The charge dissociation in water-ethanol mixtures is expected to be lower than in pure water. Our present data suggest the dissociation on the level of ca. $20 \%$ of that in water, which means the actual charges are below the saturation values for these particles at some volume fractions. Finally, we should note that the non-monotonic behavior of $\tildeZeff$ is characteristic for the counterion-dominated screening regime. Our numerical estimations show that already amounts of salt comparable to the effective counterion concentration, ca. $1 \mu \text{M}$ in our systems, might lead to disappearance of the minimum in the effective interaction and effective charge as function of colloidal volume fraction.

In summary, we have studied the variation of the colloidal structure factor in charged deionized colloidal dispersions and demonstrated that its density dependence corresponds to a non-monotonic variation of the mean interparticle interaction and particle effective charge, with a minimum at about a volume fraction of $1 \%$. Our results cover a wider range of particle concentrations than in any previous study thus giving an excellent benchmark for models of electrostatic screening in colloidal dispersions. Furthermore, our findings are confirmed by computer simulations in the counterion dominated regime of both primitive and one-component models.

\begin{acknowledgments}

Part of the experiments were performed at the Swiss spallation neutron source SINQ, Paul Scherrer Institute, Villigen, Switzerland. We are grateful for the neutron beam time and help of our local contact person J. Kohlbrecher. We gratefully acknowledge financial support from the Swiss National Science foundation. L. F. R. O. and R. C. P. thank Conacyt-Mexico for financial support (grants 51669 and 46373).

\end{acknowledgments}

\end{document}